# Agent-Based Product Configuration: towards Generalized Consensus Seeking

Benoît Beroule[1], Alain-Jérôme Fougères[1,2] and Egon Ostrosi[1]

[1] University of Technology of Belfort-Montbéliard, Laboratory IRTES-M3M
90010 Belfort – FRANCE
{benoit.beroule, alain-jerome.fougeres, egon.ostrosi}@utbm.fr

[2] ESTA, School of Business & Engineering
90004 Belfort – FRANCE
ajfougeres@esta-Belfort.fr

**Abstract**
This paper will present an evolution of a fuzzy agent based platform which performed products configuration. As a first step, we used the notion of consensus to establish robust results at the end of the configuration process. We implemented the concept of "generalized consensus" which implied the consideration of consensuses from the beginning, in this way robust data are treated during the entire process and the final result enables the designer to distinguish the robust components and flexible ones in a set of configurations.

## 1. Introduction

The collaborative and distributive design concept raises several issues [1, 2]. Design software must take into account numerous parameters which all have to play a more or less important role in the final result, furthermore much information will transit between the different entities of software to communicate the intermediate outcomes and ideally update parameters if the user wishes to change one of them during the process.

To best achieve these objectives we have chosen an agent-based approach which will efficiently process the abundant information and will be easy to use [3, 4, 5].

In this article, we will focus on a product configuration platform named FAPIC (*Fuzzy Agent for Product Integrated Configuration*) which implements a collaborative and distributive approach based on agents [6, 7]. To be as close as possible to the real situation, the FAPIC platform agents deal permanently with fuzzy values [8, 9]. Such values allow a precise parameterization and consequently optimal results.

To improve the results the platform provides, the latter implements the concept of "consensuses" [10, 11]. In concrete terms, the agents are grouped to form clusters of similar agents and these clusters are considered as new agents in the process. We expect this approach will create groups of optimal similar product configurations which will enable robust parts of the product configuration to be identified and creates a range of products or just chosen among several proposals.

This paper is organized as follows: In the second part, the agent-based product configuration process will be detailed step by step. Then in the third part, the concept of "consensus" and its purpose will be presented. Moreover in the fourth part, the new concept of "generalized consensus" and its implementation in the platform will be introduced. Furthermore in the fifth part, the example of the configuration of an aerial conveyor will be detailed by using several approaches previously described in detail. Finally in the sixth part a comparison between the different approaches will be made. The last section, the conclusion shows interest in the proposed approach.

## 2. Agent-Based Product Configuration

2.1 Configuration Process

The FAPIC platform is an agent-based software design to make product configuration. Each agent represents information about the product separated into several categories we call communities [8].

The product configuration process performed by the FAPIC platform is divided into four steps (Figure 1) [9].

First, fuzzy relationships are built between agents from the same communities then between the communities themselves based on the data provided by the user. The purpose of this step is to link the agents and create a network which will be used to compute the coming result.

In the second step, the solution agents are evaluated. A rating is assigned to each of them depending on their consistency with the requirements imposed by the customer, the constraints determined by the experts and the functions the product must perform. This rating is a fuzzy value determined by the relationships created during the first step; the higher the value, the more compatible the solution.

During the third step, each solution agent determines the optimal configuration for the product concerned based on its local point of view. Then the configurations are evaluated considering all the points viewed in the previous steps.

Finally, in the fourth step, consensuses of optimum configurations are created by grouping them according to their resemblance. The purpose of this step is to provide ranges of products instead of just a sole configuration.

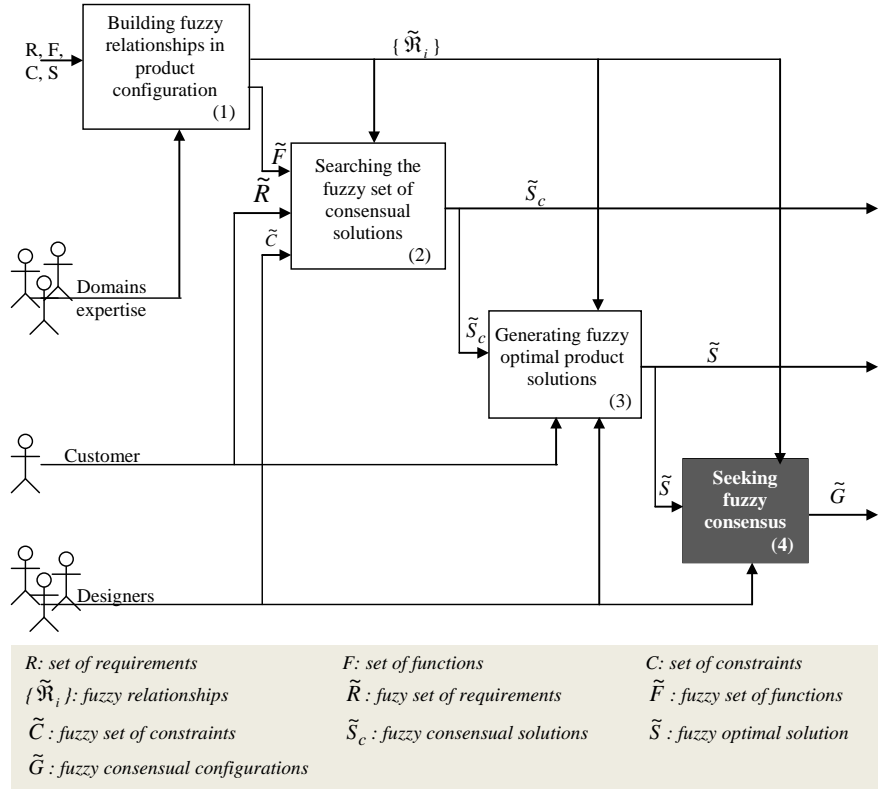

Fig. 1 Fuzzy product configuration approach

## 2.2 Configuration Process

The FAPIC platform is a fuzzy agent-based system for the distributed and collaborative configuration of products [12]. Thus we defined a fuzzy agent-based system $\tilde{M}_\alpha$ by Eq. (1):

$$\tilde{M}_\alpha = <\tilde{A}, \tilde{I}, \tilde{P}, \tilde{O}> \qquad (1)$$

Where $\tilde{A}$ is the fuzzy set of agents, $\tilde{I}$ is the fuzzy set of interactions defined in $\tilde{M}_\alpha$, $\tilde{P}$ is the fuzzy set of roles that agents of $\tilde{A}$ can play, and $\tilde{O}$ is the fuzzy set of organizations defined for fuzzy agent of $\tilde{A}$.

A fuzzy agent-based system $\tilde{M}_\alpha$ being defined, we can now define a fuzzy agent $\tilde{\alpha}_i$ of $\tilde{A}$ by Eq. (2):

$$\tilde{\alpha}_i = <\Phi_{\tilde{\Pi}(\tilde{\alpha}_i)}, \Phi_{\tilde{\Delta}(\tilde{\alpha}_i)}, \Phi_{\tilde{\Gamma}(\tilde{\alpha}_i)}, \tilde{K}_{\tilde{\alpha}_i}>. \qquad (2)$$

Where $\Phi_{\tilde{\Pi}(\tilde{\alpha}_i)}, \Phi_{\tilde{\Delta}(\tilde{\alpha}_i)}, \Phi_{\tilde{\Gamma}(\tilde{\alpha}_i)}, \tilde{K}_{\tilde{\alpha}_i}$ are respectively: the function of fuzzy observation of fuzzy the agent $\tilde{\alpha}_i$, the function of fuzzy decision of $\tilde{\alpha}_i$, the function of fuzzy action of $\tilde{\alpha}_i$, and the fuzzy knowledge of $\tilde{\alpha}_i$.

In concrete terms, the platform consists of several communities (requirements, functions, solution and one or more constraints domains) and each community contains agents which represent elements of the community (Figure 2.a).

The agents communicate by sending and receiving messages to and from other agents from linked communities to establish local results.

As shown in the figure 2 b), the process is basically divided into four parts:

1) Interactions between each $r_i$ and $f_j$ which informed the other agents of $R$ and $F$.
2) Interactions between each $f_j$ and $s_k$ which informed the other agents of $F$ and $S$
3) Interactions between each $c_l$ and $s_k$ which informed the other agents of $C$ and $S$
4) Interaction between each $s_k$ to establish the new consensual results.

Moreover the use of agents allows the platform to be dynamic because the agents are constantly running. When they are not computing calculus, they are waiting for messages from other agents. Consequently, any users (customer and/or experts) may send new data at any time. If data are sent before step 4, thus after the computation of the final result, the latter is recalculated. If data are sent during the process, the former data are updated and the final result will be changed according to the new information sent, directly without displaying the out of date result based on the initial data [13].

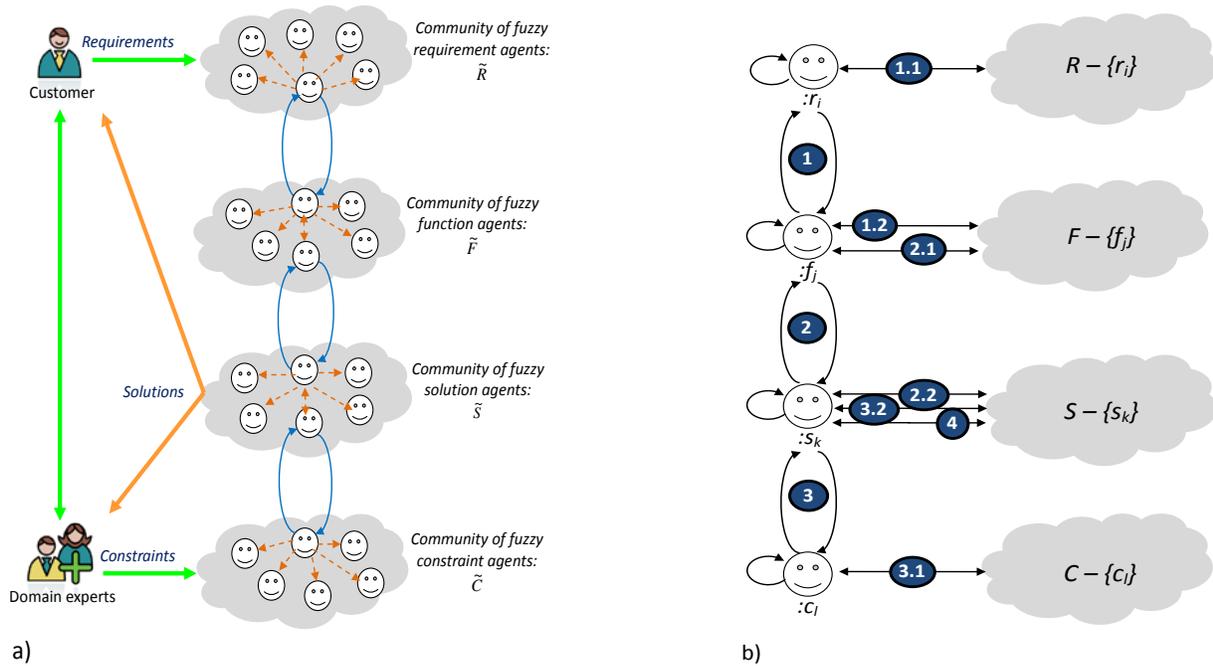

Fig. 2 a) Agent-based architecture of FAPIC platform divided in four communities of agents, and b) Illustration of interactions between fuzzy agents during Phase 2 of the product configuration (*Searching the fuzzy set of consensual solution agents*).

## 3. Configuration Consensus Seeking

3.1 Consensus Seeking

The fourth step of the configuration process consists of creating consensuses of agents between the solution and configuration communities. The purpose of this step is to create not just a sole optimal configuration but a set of configurations divided into groups named "consensuses".

Consensuses contain one or more agents which are quite similar regarding the constraints imposed by the different experts of domains and the customer. These solutions allow the customer to choose among several product configurations which best fit the imposed requirements.

To establish consensuses, the agents are divided into groups (initially one agent per group). Then an agent evaluates its knowledge to establish a new set of groups by merging the previous ones according to their fuzzy value before sending a message containing its evaluation to another agent which will do the same. Once each agent computes its own evaluation, new groups are created and the algorithm restarts until the formed groups are not modified anymore (This algorithm is detailed in the Table 1).

Table 1: Consensuses seeking algorithm (algorithm detailled in [7])

| Algorithm | Description |
|---|---|
| $DO\ by\ each\ agent\ s_i$ | // agents of S search for partition of fuzzy config. agents |
| $S'_r \leftarrow S_r$ | // save the current partition $S_r$ |
| $IF(receiveMsg(T_{G_r}))\ THEN$ | // if reception of a new partition $G_r$ |
| $\quad update(G_r);\ r \leftarrow Card(G_r)\ ENDIF$ | // update the groups of fuzzy configurations agents |
| $FOR\ p = 1\ to\ r$ | // iteration on r groups of fuzzy configuration agents |
| $\quad k1_{i,p} = \sum_{g_j \in G_p} \mu(s_i, g_j) / Card(G_p);$ | // compute k1 |
| $\quad k2_{i,p} = \sum_{g_j \notin G_p} (1 - \mu(s_i, g_j)) / Card(G - G_p)$ | // compute k2 |
| $\quad k_{i,p} = \alpha k1_{i,p} + (1-\alpha) k2_{i,p}$ | // compute k, $\alpha \in [0..1]$ |
| $ENDFOR$ | // end of iteration on r groups of fuzzy configuration agents |
| $k_{i,p*} = max(k_{i,1},...,k_{i,r})$ | // max $k$ founded in the group $p^*$ of fuzzy config. agents |
| $S_{p*} \leftarrow S_{p*} + \{s_i\}$ | // assign $s_i$ to the part of fuzzy solutions $S_{p*}$ |
| $IF(S_{p*} != S'_{p*})\ THEN$ | // inform only if modification |
| $\quad diffuse(s_i, G, T_{S_r},(s_i, S_{p*}));$ | // diffuse assigning to fuzzy configuration agents |
| $\quad diffuse(g_j, C_{p*}, T_{G_r},(g_j, G_{p*}))$ | // diffuse assigning to the fuzzy consensus cluster agents $C_{p*}$ |
| $ENDIF$ | // end of diffusion of changes |
| $WHILE(receiveMsg(T_{G_r}))$ | // while information of modifications is sent by agents of G |

### 3.2 Implementation on the FAPIC platform

In the FAPIC platform, the consensus seeking algorithm is used to create consensuses between the solutions and the configurations agents (Figure 3).

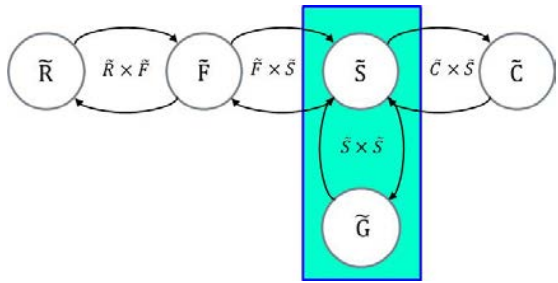

Fig. 3 Process of configuration consensus seeking on FAPIC platform

To best present the results, they are shown in a matrix which represents the fuzzy compatibility between solutions and configurations. The algorithm groups the highest values on blocks spread on the matrix diagonal. These blocks represent the consensuses; they are made up of agents which share sufficient characteristics to be considered as a range of products.

It can be necessary to rearrange some consensuses (for instance if they contain several low values), by excluding one or more configurations and obtaining a robust set of results which matches as closely as possible all the constraints and requirements.

Figure 4 shows the consensuses created by the configuration of an office chair. The solutions are divided into four groups: seat (*S1* to *S5*), back (*S6* to *S10*), armrest (*S11* to *S15*) and stand (*S16* to *S20*). Each blue rectangle highlights a set of configurations made up of four solutions (one from each groups) which create a consensus.

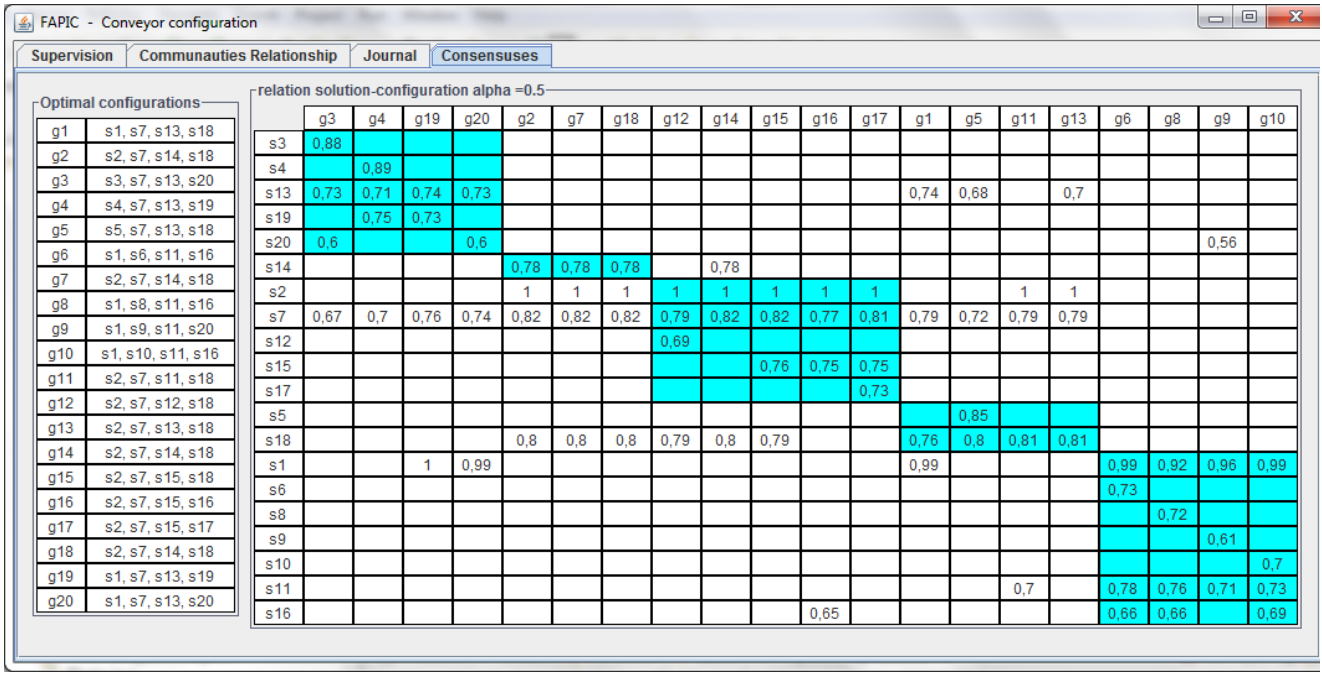

Fig. 4  Visualization of the consensuses on FAPIC platform (*Note that the matrix representation of consensus is an artificial representation, since the consensus knowledge is distributed over all agents*).

## 4. Generalized Consensus Seeking

4.1 Presentation

The next evolution of the platform should provide more robust results by introducing the concept of "generalized consensus".

The principle is to use the consensus seeking algorithm to create consensuses from each community before starting the main process, and then consider them as new super-agents containing elementary agents. The process can be run considering these super-agents which will be treated as standard agents.

In these conditions, the consensuses will be considered from the beginning of the process to obtain more robust results.

4.2 Implementation on the FAPIC platform

To apply the new concept of "generalized consensus" to the FAPIC platform, the consensuses are created directly inside the requirements, functions and constraints communities before the first step of the product configuration process. The characteristics of the new created super-agents are calculated averaging those containing elementary agents (Figure 5).

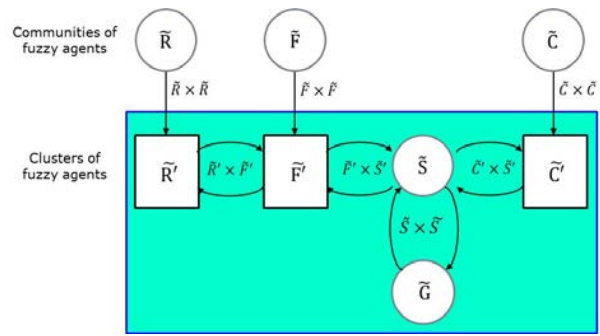

Fig. 5  Process of generalized consensus seeking on FAPIC platform

These new Agents are identical to the previous ones from a computer science point of view, so the platform may run the process normally with these agents to obtain an optimal result (step 3) composed of a set of consensuses.

To create consensuses inside a community, the relations between the agents of this community must be defined by the operators concerned (the customer or the experts) by filling the relative matrix directly on the user interface.

If these relations are not defined properly, the consensuses cannot be created and the platform will use the elementary agents during the configuration process.

## 5. Example – A Conveyor

5.1 Presentation

To illustrate the concept of "generalized consensus", the process will be detailed on a concrete example. The product to configure is an aerial conveyor used to transport heavy loads in a factory.

In this case, the platform will be parameterized with function, requirement and solution communities (the constraints are not taken into account).

The requirements are defined by the customer (see Table 2). Then the set of functions are established (see Table 5) thanks to the functional diagram (see Figure 8), with the following notations:

- CS : Control Signal
- EE : Electrical Energy
- ME : Mechanical Energy

Finally, several solutions are proposed and evaluated to perform each function (see Table 3). The evaluation is a fuzzy value which represents the effectiveness of the solution concerned. Then the set of solutions may be deduced (see Table 4).

In this case, the requirement and the function communities are defined with internal relations which enable the creation of consensuses to apply the generalized consensus. No constraints are taken into account and the solution community does not need to be divided into consensuses.

5.2 Results without the generalized consensus

If we compute the optimal configuration without using the concept of generalized consensus, we obtain the following optimal result (Figure 6).

| S1  | S2  | S3  | S5  | S7  | S9  | S10 |
|-----|-----|-----|-----|-----|-----|-----|
| S12 | S15 | S18 | S20 | S23 | S25 | S28 |

Table 2: Convoyor's requirements

Fig. 6  Simple optimal result

This result is considered as optimal but is accompanied by a set of configuration consensuses. Each consensus contains a set of configurations which are optimal according to the local point of view of at least one solution agent, these configurations form a group of products which share similar characteristics.

5.3 Results with the generalized consensus

If we compute the optimal configurations using the concept of generalized consensus, the FAPIC platform will determine consensuses from the function and the requirement communities. These new agents will be used in the process to obtain the following results finally (Figure 7).

| S1  | S2  | S3  | S5  | S7  | S9  | S10 |
|-----|-----|-----|-----|-----|-----|-----|
| <u>S12</u> | S15 | S18 | S20 | S23 | S25 | <u>S28</u> |

| S1  | S2  | S3  | S5  | S7  | S9  | S10 |
|-----|-----|-----|-----|-----|-----|-----|
| <u>S13</u> | S15 | S18 | S20 | S23 | S25 | <u>S29</u> |

| S1  | S2  | S3  | S5  | S7  | S9  | S10 |
|-----|-----|-----|-----|-----|-----|-----|
| <u>S12</u> | S15 | S18 | S20 | S23 | S25 | <u>S29</u> |

| S1  | S2  | S3  | S5  | S7  | S9  | S10 |
|-----|-----|-----|-----|-----|-----|-----|
| <u>S13</u> | S15 | S18 | S20 | S23 | S25 | <u>S28</u> |

Fig. 7  Optimal results with generalized consensus

We may notice that the former configuration determined without using the generalized consensus is present, but this time, there are four optimal results instead of just one. In this case only the optimal results are displayed and they already highlight a configuration consensus.

The base of each configuration found is the same and the difference is in the *S12* and *S13* solutions on the one hand then the *S28* and *S29* on the other hand in concrete terms, these results suggest that the function "import support + load" may be realized by a pivot (*S12*) or a screw (*S13*), and the function "transmit load + support on rail" may be realized thanks to welding (*S28*) or bending (*S29*). The other solutions form a robust part of the conveyor configuration.

| R1 | Easy to assemble | R13 | Environment resistant |
|---|---|---|---|
| R2 | Easy accessible pieces | R14 | Easily washable |
| R3 | Quick to install | R15 | Long life time |
| R4 | Easy to dismantle | R16 | Ad. Different speeds |
| R5 | Minimize work | R17 | Not too cumbersome |
| R6 | Min. rails degradation | R18 | Allow CAD |
| R7 | Adaptable to rails | R19 | Elect. motor admission |
| R8 | Ad. different rails | R20 | Impermeability |
| R9 | Ad. different supports | R21 | High temper. resistant |
| R10 | Respect security norms | R22 | Use client pieces |
| R11 | Minimize noise | R23 | Use easy manuf. pieces |
| R12 | Support heavy loads | R24 | Minimize price |

| S1 | Defined by customer | S16 | Hydrostatic contact |
|---|---|---|---|
| S2 | Defined by customer | S17 | Mecha. contact by sides |
| S3 | Electrical wires | S18 | 4 tensioners |
| S4 | Direct contact | S19 | 3 tensioners |
| S5 | Prog. Logic controller | S20 | Adhesion tensioner/rail |
| S6 | Contactor | S21 | Pinion/rack |
| S7 | Prog. Logic controller | S22 | Caterpillar/rail |
| S8 | Potentiometer | S23 | Cable and kook |
| S9 | Defined by customer | S24 | Metallic cable |
| S10 | Reducer | S25 | Pivot |
| S11 | Belt | S26 | Screw |
| S12 | Pivot | S27 | Mandrel |
| S13 | Screw | S28 | Welded |
| S14 | Mandrel | S29 | Bend |
| S15 | Mechanical contact | S30 | |

Table 3: Convetor's solutions

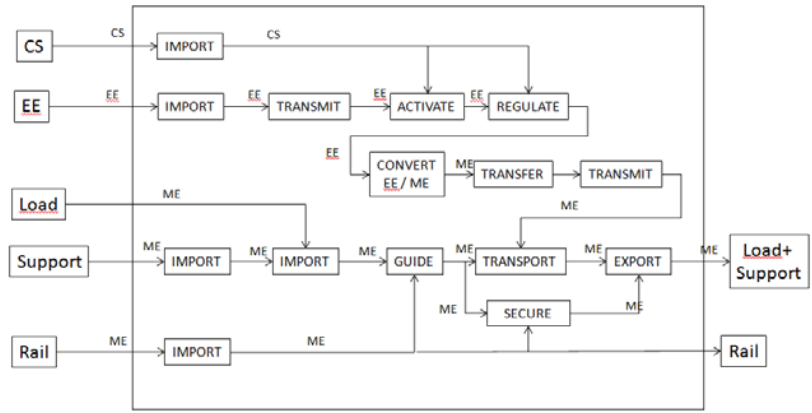

Fig. 8 Conveyor's functional diagram

Table 4: Conveyor's solution evaluation

| Sub functions \ Solutions | | S1 | S2 | S3 |
|---|---|---|---|---|
| 1 | Import CS | Defined by the customer | | 1 |
| 2 | Import EE | Defined by the customer | | 1 |
| 3 | Transmit EE | Electrical wires 0,9 | Direct contact 0,6 | |
| 4 | Activate EE | Industrial programmable logic controller 0,9 | Contactor 0,6 | |
| 5 | Régulate EE | Industrial programmable logic controller 0,9 | Potentiometer 0,6 | |
| 6 | Convert EE/ME | Defined by the customer | | 1 |
| 7 | Transmit ME | Reducer 0,9 | Belt 0,6 | |
| 8 | Import support + load | Pivot 0,9 | Vise 0,6 | Mandrel 0,6 |
| 9 | Import rail | Mecanical contact 0,9 | Hydrostatical contact 0,6 | Mecanical contact by sides 0,6 |
| 10 | Guide Load+Support on rail | 4 tensioners 0,9 | 3 tensioners 0,6 | |
| 11 | Transport Load+Support on rail | Adhesion tensioner/rail 0,9 | Pinion/rack 0,6 | caterpillar/Rail 0,6 |
| 12 | Secure Load+Support on rail | Cable + Hook 0,9 | Metalical cable 0,6 | |
| 13 | Export Load+Support | Pivot 0,9 | Vise 0,6 | Mandrel 0,6 |
| 14 | Transmit Load+Support on rail | Welded 0,9 | bend 0,6 | |

Table 5: Conveyor's Functions

| F1 | Import CS |
|---|---|
| F2 | Import EE |
| F3 | Transmit EE |
| F4 | Activate EE |
| F5 | Regulate EE |
| F6 | Convert EE/ME |
| F7 | Transmit ME |
| F8 | Import support + load |
| F9 | Guide support + load |
| F10 | Import support + load |
| F11 | Transport support + load |
| F12 | Secure support + load |
| F13 | Export support + load |
| F14 | Transmit support + load |

## 7. Conclusions

The purpose of the implementation of the generalized consensus concept was to obtain more robust results. With this approach, the notion of consensuses is present at each step of the configuration process and we may observe that the change affects the process because the latter takes this notion directly into account and provides a group of similar results which will be used to create a range of products without further processing.

The next step is to use the concepts and models developed in the FAPIC platform to apply it to the "city organization". We consider the implementation may be duplicated with several arrangements to respond to numerous problems which will imply considering the contribution of a large amount of entities.

**Benoît Beroule** received the B.S. degree from the department of Computer Science of University of Technology of Belfort-Montbéliard (UTBM). He is currently pursuing the Ph.D. degree in the Laboratory IRTES-SeT (Systems and Transports) of UTBM.

**Alain-Jérôme Fougères** holds a PhD in Artificial Intelligence from the University of Technology of Compiègne - France (1997). He is a Professor of Computer Science at ESTA and he conducts his research at ESTA-Lab. He is also associated with the Laboratory IRTES-M3M of University of Technology of Belfort-Montbéliard. Initially, his areas of interests and his scientific contributions were: 1) assistance to the redaction of formal specifications, where problems focused on the natural language processing, the knowledge representation, and the techniques of formal specifications; and 2) design of agent-based systems, in particular architecture, interactions, communication, and cooperation problems. Over the last ten years, his research has focused on agent-based mediation for cooperative work.

**Egon Ostrosi** holds a Ph.D. from the University "Louis Pasteur" of Strasbourg (ULP), and is currently Associate Professor in the Laboratory of Mecatronics3M (M3M) of IRTES at the University of Technology of Belfort – Montbéliard. His current research concerns integrated mechanical product design and modeling, knowledge engineering and its application in mechanical design, and concurrent engineering. He has been actively involved with the development of methods for concurrent product design, CAD product modeling, product configuration, and collaborative and distributed design process.